%% file: main.tex
\documentclass[10pt,twocolumn,a4paper]{article}

\usepackage[T1]{fontenc}
\usepackage[utf8]{inputenc}
\usepackage[english]{babel}
\usepackage{lmodern}
\usepackage[a4paper,top=1.28cm,bottom=1.34cm,left=1.31cm,right=1.31cm,columnsep=0.54cm]{geometry}
\usepackage{graphicx}
\usepackage{booktabs}
\usepackage{array}
\usepackage{tabularx}
\usepackage{microtype}
\usepackage{enumitem}
\usepackage{xurl}
\usepackage{url}
\usepackage[hidelinks]{hyperref}
\usepackage{xcolor}
\usepackage{tikz}
\usepackage{caption}
\usepackage{flushend}
\usepackage[numbers,sort&compress]{natbib}
\usepackage[hidelinks]{hyperref}
\usepackage{tikz}
\usepackage{eso-pic}
\hypersetup{
  pdftitle={RO-LiDAR GeoQuickView: A Web Platform for Exploring Public LiDAR-Derived Elevation Data in Romania},
  pdfauthor={Alexandru Hegyi}
}

\setlist[itemize]{leftmargin=*,topsep=2pt,itemsep=0pt,parsep=0pt}
\setlist[enumerate]{leftmargin=*,topsep=2pt,itemsep=0pt,parsep=0pt}

\newcommand{\platformname}{RO-LiDAR GeoQuickView}
\newcommand{\authorname}{Alexandru Hegyi}
\newcommand{\authoremail}{alexandru.hegyi@geo.uio.no} 

\newcommand{\targetcrs}{EPSG:3844 / Stereo 70}

\newif\ifdownloadlive
\downloadlivetrue

\newif\ifzoneBlive
\zoneBlivefalse

\ifdownloadlive
  \newcommand{\downloadstatus}{The public interface includes a clickable spatial index for direct retrieval of linked raster packages.}
  \newcommand{\downloadabstract}{provides a spatial index for direct raster retrieval}
\else
  \newcommand{\downloadstatus}{A clickable spatial index for direct retrieval of linked raster packages has been prepared and is being integrated into the public interface.}
  \newcommand{\downloadabstract}{is being extended with a spatial index for direct raster retrieval}
\fi

\ifzoneBlive
  \newcommand{\zoneBstatus}{The high-resolution LAKI III component includes both Zone A and Zone B. Zone B adds Suceava, Neam\c{t}, Bac\u{a}u, and Vrancea counties.}
  \newcommand{\zoneBroadmap}{Following the integration of LAKI III Zone B, development will focus on refined metadata, contextual layers, and scalable open delivery.}
\else
  \newcommand{\zoneBstatus}{LAKI III Zone B, comprising Suceava, Neam\c{t}, Bac\u{a}u, and Vrancea counties, is the next scheduled high-resolution extension. It will be integrated after the public products become available and pass the same harmonization and quality-control workflow.}
  \newcommand{\zoneBroadmap}{The immediate priority is the processing and integration of LAKI III Zone B for Suceava, Neam\c{t}, Bac\u{a}u, and Vrancea counties after the public products become available.}
\fi

\title{%
{\bfseries\LARGE RO-LiDAR GeoQuickView: A Web Platform for Exploring Public LiDAR-Derived Elevation Data in Romania}%
\thanks{%
\textbf{Platform access.}
RO-LiDAR GeoQuickView is publicly available through the interactive web application:
\href{https://experience.arcgis.com/experience/a48e1284ad9f4b4cb8bb75bd96f57d49/page/webapp}
{\textbf{Access the RO-LiDAR GeoQuickView platform}}.
}%
}
\author{\authorname\\[-1mm]
\small Department of Geosciences, University of Oslo, Norway\\[-1mm]
\small Correspondence: \texttt{\authoremail}}

\date{}

\begin{document}
\maketitle
\vspace{-0.72cm}

\begin{abstract}
\noindent Public elevation data can support landscape research, environmental interpretation, planning, education, and public engagement, but practical reuse is often limited by fragmented delivery and specialist processing requirements. This paper presents \platformname, an independent, voluntary, and non-commercial Web-GIS initiative for exploring and reusing publicly accessible elevation data in Romania. The platform integrates LiDAR-derived digital terrain models (DTMs) and complementary elevation models of different resolutions, publishes standardized hillshade visualizations for immediate browser access, supports participatory landscape documentation, and \downloadabstract. Its most detailed currently integrated component is the 0.5~m LAKI III Zone A DTM coverage for Cara\c{s}-Severin, Gorj, Mehedin\c{t}i, and Dolj counties. \zoneBstatus{} The platform also incorporates LAKI II and additional public altimetric sources through a source-aware processing workflow that accommodates different acquisition units, including one-kilometre cells and larger raster blocks. The paper documents the data architecture, harmonization steps, quality-control procedures, access modes, application range, and limitations. The platform is conceived as an accessibility layer over public data infrastructures: it supports rapid discovery and preliminary interpretation, but it does not replace official products, specialist modelling, expert review, or field verification.
\end{abstract}

\vspace{0.5mm}

\section{Introduction}
Airborne laser scanning (ALS), commonly referred to as airborne LiDAR, is an active remote-sensing method that estimates distances by recording the travel time of emitted laser pulses and their returns \citep{wehrlohr1999}. Combined with positioning and orientation information, the measurements generate georeferenced three-dimensional point clouds that describe terrain and surface structure \citep{wehrlohr1999}. Ground-classified points can be interpolated into a digital terrain model (DTM), whereas a digital surface model (DSM) retains upper surfaces such as vegetation, buildings, and other objects \citep{sitholevosselman2004}. Because ground filtering is an interpretative processing step, DTM quality depends on terrain complexity, vegetation, point density, and the selected classification method \citep{sitholevosselman2004}.

Raster derivatives make large elevation datasets easier to inspect visually. Analytical hillshade simulates illumination over a terrain model and can improve the perception of slopes, scarps, terraces, channels, and other relief features \citep{kokaljhesse2017}. A single hillshade is not exhaustive because the apparent visibility of a feature depends partly on its orientation relative to the virtual light source \citep{kokaljhesse2017}. Complementary products, including multidirectional shading, sky-view factor, openness, slope, and local-relief representations, can reduce directional bias or highlight different scales of topographic variation \citep{kokaljhesse2017,zaksek2011}.

The application range of detailed terrain information is broad. LiDAR-derived elevation models are widely used in landslide detection, characterization, modelling, and monitoring \citep{jaboyedoff2012}. Accurate terrain representation is important for hydrological modelling and the analysis of drainage pathways \citep{hollaus2005}. ALS point clouds and derivatives have become important resources for forest inventory and ecosystem assessment \citep{maltamo2014}. Dedicated LiDAR systems and derived metrics are increasingly used in precision-agriculture research, including crop characterization and field-management applications \citep{farhan2024}. Public elevation infrastructures can therefore serve research, environmental management, territorial planning, education, and local landscape knowledge.

Romania has an increasingly valuable but heterogeneous collection of public elevation datasets. The National Agency for Cadastre and Land Registration (ANCPI) coordinated LAKI II and LAKI III, projects intended to deliver elevation and related geospatial products for large areas of the country \citep{ancpiLaki2,ancpiLaki3}. LAKI III targets approximately 50,000~km$^2$ and identifies DTM and DSM production through airborne LiDAR scanning as a basis for environmental monitoring, water-resource management, forest-resource assessment, natural-risk analysis, agriculture, urban planning, transport and communication infrastructure, archaeology, heritage protection, and improved geographic-data distribution \citep{ancpiLaki3}.

Publishing elevation data is necessary but does not automatically make them easy to inspect or reuse. Large datasets distributed through multiple source products still require discovery, indexing, validation, GIS harmonization, quality control, and web optimization. \platformname{} was developed to reduce this practical gap. It does not replace primary providers or create a new survey. Instead, it connects public elevation products to browser visualization, spatial data discovery, raster retrieval, metadata, and optional structured observations.

\begin{figure}[t]
  \centering
  \includegraphics[width=\columnwidth]{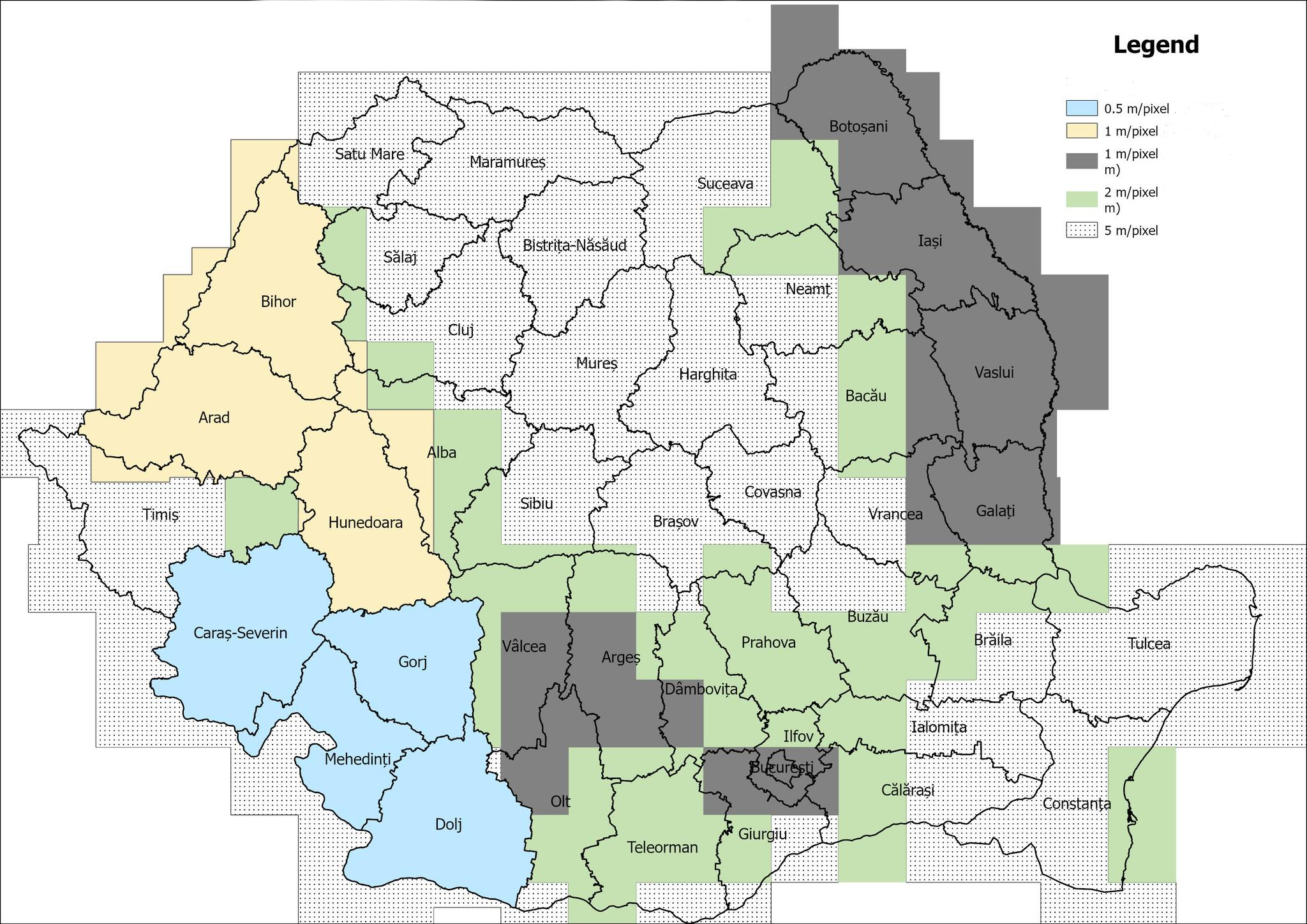}
  \caption{Spatial distribution and nominal ground resolution of the elevation datasets integrated into RO-LiDAR GeoQuickView. The map illustrates the heterogeneous, multi-source character of the available altimetric data across Romania.}
  \label{fig:platform}
\end{figure}

\section{From GeoQuickView to a web accessibility layer}
The conceptual starting point for \platformname{} was GeoQuickView, a desktop application designed to simplify LiDAR-derived raster visualization for users with limited GIS experience \citep{hegyi2024}. The web platform extends the same accessibility principle from local raster inspection to large-area online exploration.

The current implementation uses ArcGIS Experience Builder, a configurable environment for assembling responsive web experiences from maps, data layers, and interactive widgets \citep{esriExperience}. Four design principles guide the application:
\begin{enumerate}
  \item \textbf{Immediate legibility.} Users should be able to inspect terrain through standardized visualizations without installing software or preprocessing source rasters.
  \item \textbf{Progressive complexity.} Browser exploration should remain simple, while advanced users retain a pathway to raster retrieval and independent GIS processing.
  \item \textbf{Transparent limitations.} Hillshade is an analytical visualization, not a photograph and not a definitive interpretation. Dataset resolution, acquisition history, and artefacts require documentation \citep{kokaljhesse2017}.
  \item \textbf{Cross-sector reuse.} The same terrain layer can support different questions in geomorphology, hydrology, forestry, agriculture, planning, education, cultural-landscape research, and local documentation.
\end{enumerate}

The result is a general-purpose terrain-exploration and data-discovery interface. Participatory reporting is an additional component rather than the sole destination of the workflow.

\section{Public datasets and integrated coverage}
\subsection{LAKI II and western Romanian datasets}
LAKI II was developed under the project ``Geographic information for the environment, climate change and EU integration''. Its objectives included LiDAR-derived DTM and DSM products and associated geospatial information for approximately 50,000~km$^2$ in Arad, Bihor, Hunedoara, Alba, Mure\c{s}, and Harghita counties \citep{ancpiLaki2Objectives,cncLaki2}. Publicly distributed products available for reuse form an important western high-resolution component of \platformname{}.

Freely accessible LAKI II DTMs have already supported studies of the Dacian Kingdom landscape. Pe\c{t}an and Hegyi \citep{petanhegyi2023} emphasized the interpretative value of freely available LiDAR-derived DTMs; Pe\c{t}an \citep{petan2023terraces} discussed settlements on terraces in the central area of the Dacian Kingdom; and a recent synthesis examines the terraced mountains as a large-scale transformed Late Iron Age landscape \citep{petan2026terraces}. These studies are relevant precedents for accessible public data, while archaeology remains one application within the wider platform.

\subsection{LAKI III Zones A and B}
LAKI III continues DTM and DSM production through airborne LiDAR scanning. ANCPI defines two project zones: Zone A comprises Cara\c{s}-Severin, Gorj, Mehedin\c{t}i, and Dolj counties, while Zone B comprises Suceava, Neam\c{t}, Bac\u{a}u, and Vrancea counties \citep{ancpiLaki3}. The most detailed products currently integrated into \platformname{} are the 0.5~m Zone A DTMs. \zoneBstatus{}

\subsection{Complementary elevation models}
A uniform national high-resolution LiDAR survey is not yet available. To preserve regional continuity, the platform also integrates complementary public altimetric products. Depending on the region, these are predominantly 1~m or 2~m products, locally supplemented by approximately 5~m models. An extended approximately 5~m representation is useful for broad terrain exploration but should not be interpreted as equivalent to the 0.5~m or 1~m LiDAR-derived DTMs \citep{guta2026}. Table~\ref{tab:coverage} summarizes the conceptual data groups.

\begin{table*}[t]
\centering
\caption{Elevation-data groups represented in \platformname{} and their role in the platform.}
\label{tab:coverage}
\small
\begin{tabularx}{\textwidth}{@{}p{3.08cm}p{4.38cm}p{1.98cm}p{2.28cm}X@{}}
\toprule
\textbf{Data group} & \textbf{Coverage} & \textbf{Nominal resolution} & \textbf{Status} & \textbf{Function in the platform} \\
\midrule
LAKI III Zone A & Cara\c{s}-Severin, Gorj, Mehedin\c{t}i, Dolj & 0.5~m & Integrated & Highest-detail terrain visualization; one-kilometre source-cell workflow; county mosaics and hillshades \\
LAKI III Zone B & Suceava, Neam\c{t}, Bac\u{a}u, Vrancea & High resolution & Scheduled extension & Same source-aware harmonization, QC, mosaicking, and publication workflow after public release \\
LAKI II public products & Western Romanian coverage, including Arad, Bihor, Hunedoara, and part of Alba & 1~m & Integrated & High-resolution terrain exploration; larger-block acquisition; reusable research data \\
Complementary public models & Additional regions of Romania & Predominantly 1--2~m; locally about 5~m & Integrated selectively & Regional continuity and broader terrain exploration \\
Extended terrain representation & Wider national context & About 5~m & Integrated & Large-landform overview; not equivalent to fine-scale LiDAR-derived coverage \\
\bottomrule
\end{tabularx}
\end{table*}

\section{Multi-source processing and raster retrieval}
\subsection{Source-aware acquisition units}
The platform was designed for heterogeneous public products rather than for a single distribution scheme. LAKI III Zone A is distributed as numerous one-kilometre terrain-model packages. The LAKI III acquisition route therefore uses a fine grid: candidate cells are generated, validated against the public source, and retrieved only when a corresponding package is available. LAKI II and several complementary altimetric collections are acquired through larger provider units, including 40~km $\times$ 40~km source blocks in parts of the integrated workflow. Other collections use provider-specific blocks. The acquisition unit changes with the source, but the harmonization logic remains consistent.

This distinction is important for reproducibility. The platform records the source project, native resolution, acquisition unit, spatial extent, and processing status before mosaicking. Source-aware indices avoid repeated retrieval, preserve an audit trail for unavailable or problematic units, and allow later rechecks when providers publish additions or corrections.

\subsection{Standardized harmonization and quality control}
After acquisition, the products pass through the same general sequence: extraction, metadata inspection, spatial-reference checking, organization by source and resolution class, mosaicking, clipping, quality control, derivative generation, and web publication. Input products are harmonized for publication while their provenance is retained. The current web workflow uses the Romanian national Stereo 70 reference system (\targetcrs) where appropriate for local processing and index generation.

Quality control is required at several levels. The workflow inspects missing units, NoData masks, anomalous values, edge artefacts, and visible seams. Regional or county mosaics and cutlines are constructed only after the source units have been checked. A source-aware approach is preferable to treating all products as if they were identical because spatial resolution, acquisition history, native tiling, and processing lineage affect what users can interpret reliably.

The browser visualization currently uses a north-west virtual illumination direction with an azimuth of 315$^{\circ}$ and a solar-elevation angle of 35$^{\circ}$ \citep{guta2026}. This standardized rendering provides a coherent overview, but detailed local interpretation should compare alternative visualization methods because a single illumination direction can suppress features aligned with the virtual light source \citep{kokaljhesse2017}.

\subsection{Complementary access modes}
The platform separates immediate exploration from advanced reuse. The hillshade viewer supports rapid terrain inspection without requiring local preprocessing. \downloadstatus{} The spatial index is designed to let users identify relevant grid cells or source units and retrieve the corresponding raster packages for processing in GIS software. This dual-access model supports non-specialists who need an interpretable overview and advanced users who require underlying rasters for independent analysis.

\begin{figure*}[t]
  \centering
  \resizebox{0.98\textwidth}{!}{\input{figures/workflow.tex}}
  \caption{Source-aware workflow of \platformname. Acquisition units vary by provider, while the platform applies a consistent sequence of harmonization, quality control, derivative generation, and delivery.}
  \label{fig:workflow}
\end{figure*}

\begin{figure}[t]
  \centering
  \includegraphics[width=\columnwidth]{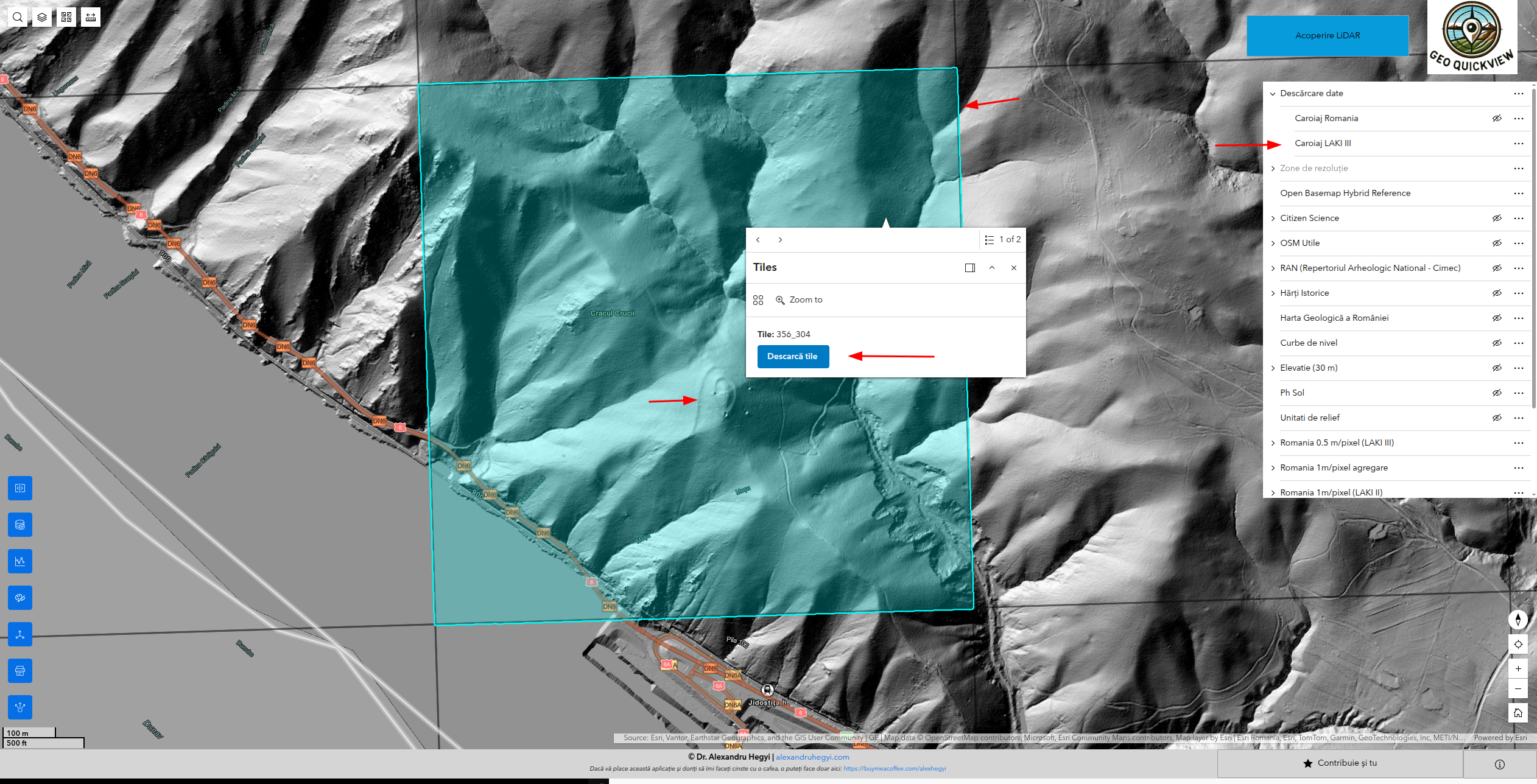}
\caption{Example of the raster-tile retrieval interface in RO-LiDAR GeoQuickView. A LAKI III grid cell is selected directly from the interactive map and highlighted in turquoise.}
  \label{fig:download}
\end{figure}

\section{Cross-sector landscape applications}
\platformname{} is intended as an exploratory interface, not as a replacement for specialist analytical systems. Table~\ref{tab:applications} distinguishes what can be inspected directly in the browser from the additional evidence needed for operational interpretation.

\begin{table*}[t]
\centering
\caption{Application range of \platformname{} and the distinction between browser-based exploration and specialist follow-up analysis.}
\label{tab:applications}
\small
\begin{tabularx}{\textwidth}{@{}p{3.15cm}p{6.15cm}X@{}}
\toprule
\textbf{Application area} & \textbf{What the platform can support directly} & \textbf{What requires additional data, modelling, or verification} \\
\midrule
Geomorphology and environmental screening & Visual inspection of scarps, terraces, ravines, sinkholes, abandoned channels, quarry surfaces, and broad terrain organization & Geological context, field mapping, process analysis, and specialist interpretation \\
Hydrology, drainage, and natural risks & Preliminary observation of drainage pathways, low-lying terrain, floodplain morphology, embankments, and possible mass-movement signatures & Hydrological or hydraulic models, rainfall and discharge data, engineering inputs, susceptibility analysis, and field validation \\
Forestry and natural resources & Terrain accessibility, roads, ravines, slope context, and landform interpretation below vegetation & Point-cloud or DSM-derived canopy metrics, inventories, ecological observations, and management data \\
Agriculture and land use & Terrain context for drainage, erosion pathways, terraces, embankments, field morphology, and land-use traces & Crop, soil, multispectral, management, and in-situ information \\
Infrastructure and territorial planning & Initial terrain screening for route corridors, slopes, settlement edges, and local constraints & Engineering design, cadastral information, utilities, geological and geotechnical data, and regulatory assessment \\
Education, fieldwork, and public engagement & Terrain literacy, comparison of landscapes, preparation of field visits, mobile orientation, and public curiosity & Guided interpretation, contextual teaching material, and appropriate safety procedures \\
Cultural landscapes, archaeology, and local memory & Preliminary observation of routeways, earthworks, terraces, platforms, mounds, historic land use, and toponyms & Specialist review, historical sources, field verification, heritage procedures, and protection of sensitive coordinates \\
Repeat-survey monitoring & Traceable source units and reproducible derivatives that can support later comparisons & Harmonized multi-temporal data, co-registration, uncertainty assessment, and change-detection methods \\
\bottomrule
\end{tabularx}
\end{table*}

\subsection{Terrain systems, water resources, and natural risks}
High-resolution DTMs can reveal the spatial organization of hillslopes, fluvial corridors, karst terrain, and mass-movement morphology \citep{jaboyedoff2012,kokaljhesse2017}. Landslide investigations use LiDAR-derived terrain models for detection, characterization, susceptibility assessment, modelling, and monitoring \citep{jaboyedoff2012}. Hydrological analysis also depends on terrain representation, particularly where flow paths, river corridors, floodplains, and embankments must be represented coherently \citep{hollaus2005}. Romanian studies illustrate this practical value: Hu\c{t}anu et al. \citep{hutanu2020} used a 0.5~m LiDAR-derived DEM with HEC-RAS modelling in the Jijia floodplain, while Cop\u{a}cean et al. \citep{copacean2025} used a 1~m DEM generated through LAKI I and LAKI II for a Cigher River flood assessment. \platformname{} does not perform operational risk modelling, but it can support preliminary screening and the identification of areas that justify more detailed analysis.

\subsection{Forestry, agriculture, and territorial planning}
In forestry and natural-resource management, terrain models help describe access routes, slopes, ravines, and landform context below vegetation. Full forest analysis normally requires point clouds, DSMs, canopy-height models, inventories, and field data; ALS is widely used for forest inventory and ecosystem assessment \citep{maltamo2014}. A Romanian Porolissum study demonstrated a particularly useful cross-sector case: LiDAR-derived models revealed both Roman frontier defensive systems and the present forest-road network, supporting integrated heritage and forest-management decisions \citep{roman2017}. In agricultural landscapes, terrain derivatives can reveal drainage patterns, embankments, terraces, erosion pathways, and field morphology. Dedicated precision-agriculture applications extend further into crop structure and management when suitable sensors and complementary datasets are available \citep{farhan2024}. For infrastructure and territorial planning, the viewer provides initial terrain screening rather than engineering design.

\subsection{Education, field orientation, and participatory documentation}
The application can support terrain literacy by connecting abstract landform concepts to familiar places. Teachers, students, field teams, and local residents can compare lowlands, valleys, hillslopes, plateaus, and mountain environments through the same interface. Participatory forms can complement raster interpretation with observations related to geomorphology, cultural landscapes, toponyms, and local memory \citep{guta2026}. Contributions should be treated as observations requiring review, not as validated interpretations.

\subsection{Cultural landscapes and archaeology}
Archaeology is a relevant but deliberately non-exclusive application. In wooded environments, ALS-derived DTMs can support the recognition of topographic earthworks that may be difficult to observe in ordinary imagery or during field visits \citep{doneus2008}. Romanian applications already span several landscape types and methodological approaches. LiDAR-derived terrain information has supported the documentation of the forested archaeological landscape and Roman frontier systems around Porolissum \citep{opreanu2014,roman2017}, the reassessment of Sarmizegetusa Regia and the wider upland organization of the Or\u{a}\c{s}tie Mountains \citep{olteanhanson2017,olteanfonte2019,olteanfonte2021}, and the investigation of the large fortified settlement at Corne\c{s}ti-Iarcuri \citep{vizireanu2018}. Other studies illustrate the use of high-resolution terrain data for the extraction of burial-mound candidates \citep{niculita2020}, the exploration of south-eastern Carpathian hill-top sites through combined methods \citep{stefanstefan2021}, and the characterization of Late Bronze Age settlements and visible ashmounds in north-eastern Romania \citep{mihupintilie2022,brasoveanu2023}.

Freely accessible LAKI II-derived DTMs have also enabled research on the central landscape of the Dacian Kingdom, including transformed mountain terrain, terraces, fortifications, and limestone-quarry areas \citep{petan2022quarries,petan2023terraces,petanhegyi2023,petan2026terraces}. Together, these studies show that accessible terrain products can stimulate research across different periods and landscape settings without redefining \platformname{} as an archaeological application. Potential archaeological features visible in the platform remain preliminary observations and require specialist review, contextual evidence, field verification, and appropriate protection of sensitive coordinates.

\subsection{Toward repeat-survey monitoring}
Repeated ALS acquisitions can support the analysis of terrain changes, including landslide activity and related geomorphological processes \citep{jaboyedoff2012}. \platformname{} is not yet a change-detection platform, but its organization around traceable source units, quality-control records, and reproducible derivatives provides a foundation for future monitoring workflows.

\section{Limitations and responsible interpretation}
The platform must be read critically. Coverage is heterogeneous: datasets differ in source, acquisition date, resolution, processing history, and intended purpose. A regional 5~m elevation model is not equivalent to a 0.5~m LiDAR-derived DTM. A DTM is also a processed representation rather than an unfiltered measurement. Ground classification can preserve vegetation artefacts or suppress genuine fine-scale terrain features, particularly in complex landscapes \citep{sitholevosselman2004}.

Hillshade is a simulated illumination product. It can emphasize some slopes, conceal others, and alter the apparent prominence of minor features \citep{kokaljhesse2017}. Natural landforms can also resemble anthropogenic structures. Landslides, karst depressions, dry valleys, ravines, fluvial scarps, geological structures, scree, forestry operations, and modern interventions may produce misleading patterns \citep{kokaljhesse2017,guta2026}. Interpretation should therefore consider terrain setting, geology, drainage, land use, vegetation, historical sources, maps, and field evidence.

Finally, the viewer is an accessibility layer rather than an official data authority. Source products remain the responsibility of their providers and must be reused according to applicable licences and attribution conditions. The platform does not replace cadastral products, precision surveying, operational risk models, forestry inventories, agricultural monitoring systems, engineering design, archaeological procedures, navigation tools, or institutional decision-making.

\section{Development roadmap}
\zoneBroadmap{} The same source-aware approach can be reused: register the public collection, identify its provider units, retrieve available products, preserve reproducible indices, inspect gaps and NoData masks, generate regional mosaics, produce web derivatives, and publish a linked discovery or retrieval layer.

A second priority is metadata refinement. Coverage maps should distinguish resolution classes, source projects, native acquisition units, and processing status clearly. Visual guides should explain the difference between point clouds, DTMs, DSMs, hillshade derivatives, downloadable rasters, and user-submitted observations. Additional visualization products can improve local interpretation without overwhelming the lightweight browser experience \citep{kokaljhesse2017,zaksek2011}.

A third direction concerns delivery architecture. The ArcGIS-based interface offers an effective public entry point, while a longer-term self-hosted architecture could use web-optimized geospatial products to preserve browser access, facilitate raster discovery, and support scalable delivery. Development will remain gradual and linked to new public data, user interest, and emerging collaborations.

\section{Conclusion}
\platformname{} demonstrates how heterogeneous public elevation products can be transformed into a practical terrain-discovery environment. Its contribution is not a new primary survey, but a source-aware workflow that connects public products to immediate browser visualization, spatial raster retrieval for local GIS reuse, metadata discovery, and optional participatory documentation. The workflow accommodates both fine one-kilometre LAKI III packages and larger LAKI II or complementary source blocks while retaining provenance and applying consistent quality control.

The platform is intentionally not restricted to a single discipline. Its broader value lies in making public terrain information more legible and reusable for geomorphology, water-resource studies, natural-risk screening, forestry, agriculture, territorial planning, education, field orientation, cultural-landscape research, local memory, and future repeat-survey monitoring. Romanian research already shows the value of detailed terrain information in flood analysis \citep{hutanu2020,copacean2025}, forest and heritage management \citep{roman2017}, and archaeological landscape interpretation \citep{petanhegyi2023,petan2026terraces}. \platformname{} generalizes the same accessibility principle across a wider range of landscape questions.

\section*{Data and software availability}
The public \platformname{} application is available online \citep{rolidarapp2026}. Source elevation datasets remain the property of their respective providers and should be reused according to applicable licences and attribution requirements. \platformname{} is an independent initiative and is not an official ANCPI application.

\section*{Acknowledgements}
The author acknowledges ANCPI and the National Centre of Cartography for publishing elevation products, as well as the open-source geospatial community whose software and documentation support reproducible terrain-data processing. The author also thanks users who contributed observations and feedback during the early public phase of the platform.

\section*{Conflict of interest}
The author declares no conflict of interest.

\end{document}

%% file: figures/workflow.tex
\begin{tikzpicture}[
  x=1cm,y=1cm,
  stage/.style={draw=black!62, rounded corners=3pt, line width=0.62pt,
                fill=black!2, text width=15.55cm, minimum height=1.05cm,
                align=left, inner xsep=7pt, inner ysep=5pt, font=\scriptsize},
  stageblue/.style={stage, draw=blue!58!black, fill=blue!3},
  stagegreen/.style={stage, draw=green!48!black, fill=green!3},
  stageorange/.style={stage, draw=orange!70!black, fill=orange!4},
  gaparrow/.style={font=\Large, text=black!72}
]
  \node[stageblue] (s1) at (0,0) {
    \textbf{1. Discover and acquire source-specific public datasets}\\[-0.4mm]
    Register source and resolution class; identify provider units; retrieve LAKI III one-kilometre cells, LAKI II larger blocks, and complementary provider-specific elevation products.
  };
  \node[gaparrow] at (0,-0.94) {$\Downarrow$};
  \node[stagegreen] (s2) at (0,-1.86) {
    \textbf{2. Harmonize, validate, and document terrain models}\\[-0.4mm]
    Check metadata and spatial reference; extract rasters; maintain source-aware indices; inspect gaps, NoData masks, seams, and anomalous values; preserve reproducible logs.
  };
  \node[gaparrow] at (0,-2.80) {$\Downarrow$};
  \node[stagegreen] (s3) at (0,-3.72) {
    \textbf{3. Produce consistent web-ready derivatives}\\[-0.4mm]
    Build regional or county mosaics and cutlines; generate standardized hillshade layers; retain the relationship between the web visualization and downloadable source rasters.
  };
  \node[gaparrow] at (0,-4.66) {$\Downarrow$};
  \node[stageorange] (s4) at (0,-5.58) {
    \textbf{4. Deliver complementary access modes}\\[-0.4mm]
    Browser terrain exploration; spatial coverage and metadata layers; raster-tile discovery and retrieval for independent GIS reuse; optional participatory reporting forms.
  };
  \node[gaparrow] at (0,-6.52) {$\Downarrow$};
  \node[stage] (s5) at (0,-7.44) {
    \textbf{5. Support cross-sector landscape applications}\\[-0.4mm]
    Geomorphology and environmental screening; hydrology and natural risks; forestry, agriculture, and planning; education and field orientation; cultural landscapes and local knowledge.
  };
\end{tikzpicture}